\documentclass[5p]{elsarticle}

\usepackage{aas_macros}
\usepackage{epsfig,amsmath,natbib}
\usepackage{amssymb}
\usepackage{amsmath}
\usepackage{dsfont}
\usepackage{hyperref}
\usepackage{color}
\usepackage{pbox}
\usepackage{booktabs}

\hypersetup{
	colorlinks=false,
	citecolor=green
}

\newcommand{\bb}[1]{\mathbf{#1}}
\newcommand{\bbh}[1]{\mathbf{\hat{#1}}}

\newcommand{\ttt}[1]{\texttt{#1}}
\newcommand{\LT}{\texttt{LensTools} }

\begin{document}

\title{Mocking the Weak Lensing universe: the LensTools python computing package}

\author[cu,bnl]{Andrea Petri}
\ead{apetri@phys.columbia.edu}

\address[cu]{Department of Physics, Columbia University, New York, NY 10027, USA}
\address[bnl]{Physics Department, Brookhaven National Laboratory, Upton, NY 11973, USA}

\date{\today}

\label{firstpage}

\begin{abstract}
We present a newly developed software package which implements a wide range of routines frequently used in Weak Gravitational Lensing (WL). With the continuously increasing size of the WL scientific community we feel that easy to use Application Program Interfaces (APIs) for common calculations are a necessity to ensure efficiency and coordination across different working groups. Coupled with existing open source codes, such as \ttt{CAMB}\citep{CAMB} and \ttt{Gadget2}\citep{Gadget2}, \LT brings together a cosmic shear simulation pipeline which, complemented with a variety of WL feature measurement tools and parameter sampling routines, provides easy access to the numerics for theoretical studies of WL as well as for experiment forecasts. Being implemented in {\sc python}\citep{python}, \LT takes full advantage of a range of state--of--the art techniques developed by the large and growing open--source software community \citep{scipy,pandas,astropy,scikit-learn,emcee}. We made the \LT code available on the Python Package Index and published its documentation on \url{http://lenstools.readthedocs.io}      
    
\end{abstract}


\maketitle


\section{Introduction}
Cosmology is entering a data driven era. After the Cosmic Microwave Background (CMB) \citep{WMAP,PlanckXVI2013} provided strong experimental evidences of cosmological theories, a variety of different probes have been proposed to unveil the secret of the cosmos. Weak Gravitational Lensing uses the correlation between image distortions of background sources by Large Scale Structure (LSS) to infer cosmological parameter values \citep{WLprimer}. Because WL probes are sensitive to late universe physics, where the density fluctuations are in the non--linear regime, quadratic features such as two--point correlation functions might miss some of the cosmological information. In the theoretical study of more complicated WL features (see for example \citep{3pcf1,bispectrum1,moments1,peaks1} for a non comprehensive list) simulation pipelines play a vital role, as in general these features cannot be predicted analytically from cosmological parameters. In this work we present a flexible, customizable and easy to deploy WL simulation pipeline that bridges the gap between simulations of shear fields, feature measurement from simulated images and cosmological parameter estimation. The paper is organized as follows: first we give an overview of the shear field simulation routines and present their runtime and memory usage benchmarks. We then outline the \LT image analysis capabilities as well as the parameter estimation routines. We finally present a summary of our work and outline our conclusions. We complement our work with some illustrative coding examples that show how to operationally use \LT for some of the former tasks.


\section{Shear simulations} 
\label{shearsim}

\subsection{Formalism}
\begin{figure*}
\begin{center}
\includegraphics[scale=0.35]{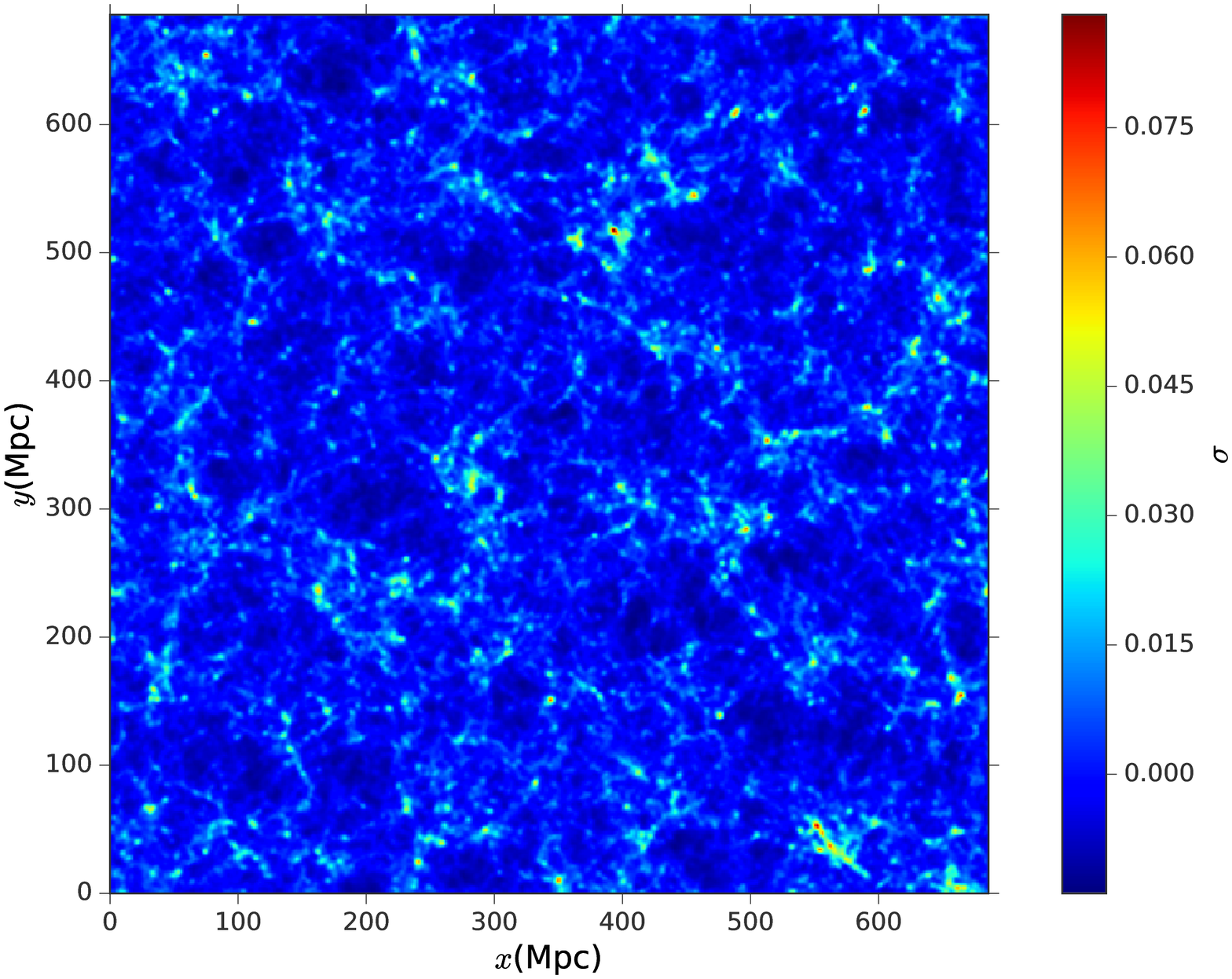}
\includegraphics[scale=0.35]{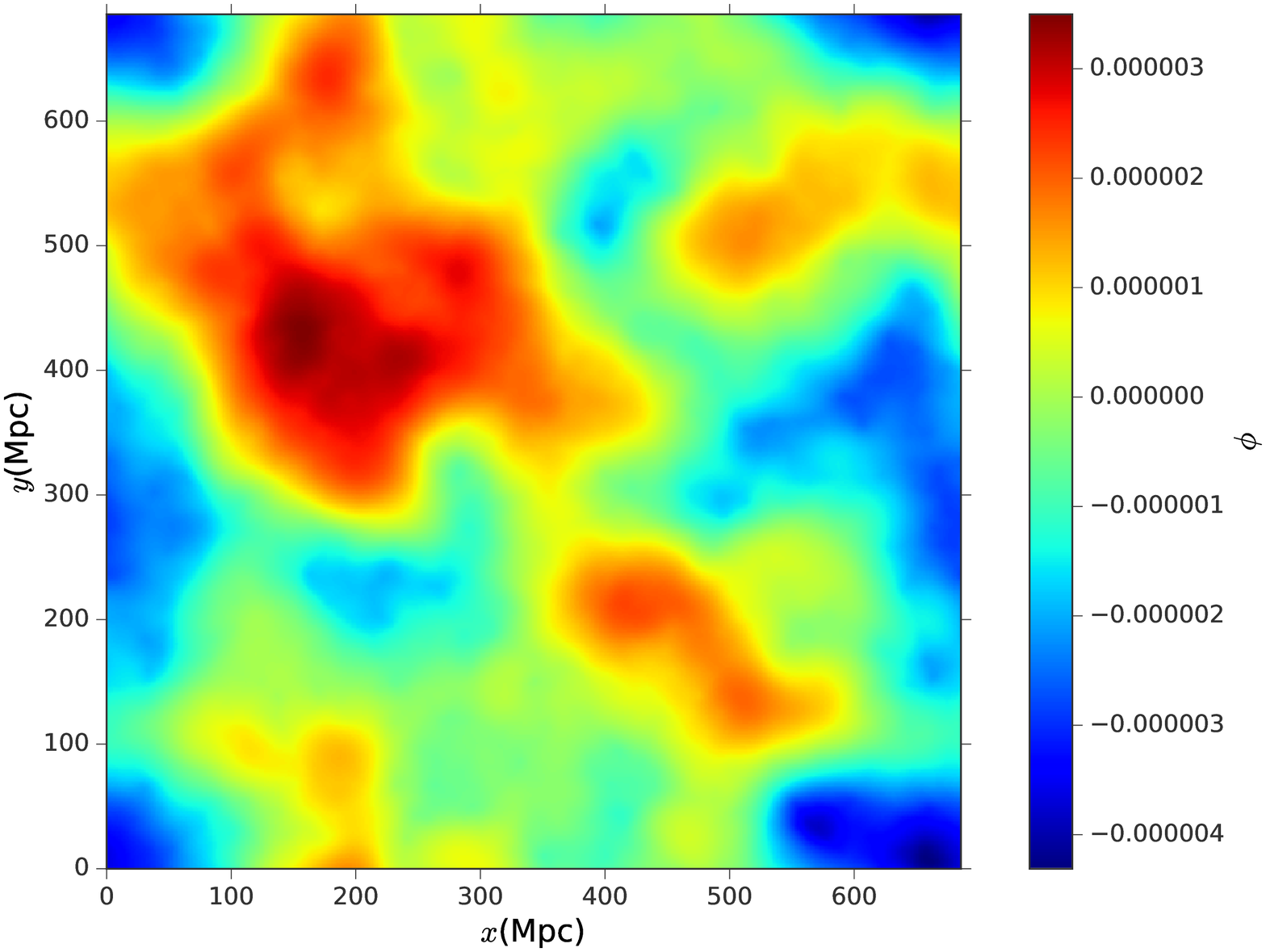}
\caption{This figure shows a lens plane, computed with a gridding procedure based on equations (\ref{surfacedensity}),(\ref{poisson}). This lens plane has been generated from a $N$--body simulation of size $L_b=480\,{\rm Mpc}/h$ with $N_p=1024^3$ particles at $z_s=2$. The plane resolution is $512\,\,$pixels per side. We show both the surface density $\sigma$ (left) and the lensing potential $\phi$ (right). Because the solution of the Poisson equation (\ref{poisson}) is computed using FFTs, there is an underlying assumption about periodic boundary conditions in the lensing potential reconstruction, which can introduce edge effects as can be seen in the corners of the potential plot. Because of these edge effects, it is a good idea to choose the field of view for the ray--tracing to be smaller than the potential field by about $\sim$ 50 pixels per side.}
\label{lensplanefig}
\end{center}
\end{figure*}

In this paragraph we give an overview of the \LT shear field simulation pipeline. This consists in a series of routines that, starting from a $w$CDM cosmological model specified by the cosmological parameters 
\begin{equation}
\bb{p}=(h,\Omega_m,\Omega_\Lambda,\Omega_b,w_0,w_a,n_s,\sigma_8)
\end{equation} 
produces random realizations of shear fields $\pmb{\gamma}(\pmb{\theta})$ in cosmology $\bb{p}$. Here $\pmb{\theta}=(\theta_x,\theta_y)$ is the angle on sky as seen from the observer. Given a background source (such as a galaxy) at redshift $z_s$, the dark matter density fluctuations $\delta(\bb{x},z)$ between the observer and the source will cause its apparent shape to be distorted due to the gravitational lensing effect, as predicted by General Relativity \citep{wlreview}. The apparent source ellipticity, assuming the unperturbed shape is a circle, can be estimated in terms of the cosmic shear $\pmb{\gamma}$ defined in equation (\ref{convsheardef}). Non circular shapes can be modeled with redshift--dependent shape noise terms \citep{wlreview}, the treatment of which goes beyond the scope of this paper. The multi--lens--plane algorithm \citep{RayTracingHartlap,Ray1,Ray2,Ray3} is a popular technique to compute light ray deflections across the path $z\in[z_s,0]$ and hence to compute the apparent source shape distortion. The mass distribution between the source and the observer is approximated as a finite set of two dimensional lenses perpendicular to the line of sight, of thickness $\Delta$ and a surface density $\sigma$ given by 

\begin{equation}
\label{surfacedensity}
\sigma(\bb{x},z) = \frac{3H_0^2\Omega_m\chi(z)}{2c^2a(z)}\int_\Delta d\chi'\delta\left(\bb{x},z(\chi')\right)
\end{equation}
where $\chi$ is the lens comoving distance and $a=1/(1+z)$ the scale factor. A light ray crossing a lens at redshift $z$ at a transverse position $\bb{x}$ will be deflected by a small angle $\pmb{\alpha}$ which can be shown to be the gradient of the 2D gravitational potential $\phi$ (see again \citep{RayTracingHartlap}) 
\begin{equation}
\label{poisson}
\nabla^2_\bb{x} \phi(\bb{x},z) = 2\sigma(\bb{x},z)
\end{equation}

\begin{equation}
\pmb{\alpha}(\bb{x},z) = \nabla_\bb{x} \phi(\bb{x},z)
\end{equation}

\begin{equation}
\bb{T}(\bb{x},z) = \nabla_\bb{x}\nabla_\bb{x}^T \phi(\bb{x},z)
\end{equation}
where we indicated $\bb{T}$ as the gradient of the deflection, which will be called distortion tensor throughout the rest of the paper. An example of a lens plane computed with the \LT pipeline according to equations (\ref{surfacedensity}),(\ref{poisson}) is shown in Figure \ref{lensplanefig}. 

The trajectory of a light ray $\bb{x}(z)$ follows the geodesic equation

\begin{equation}
\label{geodesic3D}
\frac{d^2\bb{x}}{d\chi^2} = -\frac{2}{c^2}\nabla_\bb{x_\perp}\Phi(\bb{x},z)
\end{equation}
which can be translated into a second order differential equation for the light ray angular position $\pmb{\beta}(z)=\bb{x}_\perp(z)/\chi(z)$ as seen from the observer. Following \citep{RayTracingHartlap}, the trajectory of each light ray originating at $\pmb{\beta}(0)=\pmb{\theta}$ can be calculated solving numerically a discretized version of (\ref{geodesic3D}), assuming a finite number of lenses placed at redshifts $\{z_k\}$:

\begin{equation}
\label{geodesic2D}
\pmb{\beta}_{k} = \pmb{\theta} + \sum_{i=1}^k\delta\pmb{\beta}_i 
\end{equation}

\begin{equation}
\label{deflections}
\delta\pmb{\beta}_{k+1} = (B_k-1)\delta\pmb{\beta}_k + C_k\pmb{\alpha}_k \,\, ; \,\, \delta\pmb{\beta}_0=0
\end{equation}

\begin{equation}
\label{jacobian}
\bb{A}_{k} = \mathds{1}_{2\times2} + \sum_{i=1}^k\delta\bb{A}_i 
\end{equation}

\begin{equation}
\label{sheartensorproduct}
\delta\bb{A}_{k+1} = (B_k-1)\delta\bb{A}_k + C_k\bb{T}_k\bb{A}_k \,\, ; \,\, \delta\bb{A}_0=0
\end{equation}
where $\bb{A}$ is the Jacobian matrix of the trajectory $\pmb{\beta}$ with respect to the initial light ray position $\pmb{\theta}$, $\bb{A}_k=\nabla_{\pmb{\theta}}\pmb{\beta}_k(\pmb{\theta})$. The factors $B_k,C_k$ depend on the geometry of the lens system

\begin{equation}
B_k = \frac{\chi_{k}}{\chi_{k+1}}\left(1 + \frac{\chi_{k+1}-\chi_{k}}{\chi_{k}-\chi_{k-1}}\right) \,\,\,\, ; \,\,\,\, C_k = \frac{\chi_{k}}{\chi_{k+1}} - 1 \\ \\
\end{equation} 
where we use the subscript $k$ to indicate the redshift $z_k$ of the $k$--th lens for notational simplicity. After tracing the evolution of $\bb{A}$ from the observer to the source at $z_s$, we are able to evaluate the cosmic shear $\pmb{\gamma}$ and convergence $\kappa$ at $z_s$ looking at the components of $\bb{A}$
\begin{equation}
\label{convsheardef}
\bb{A}(\pmb{\theta},z_s) = 
\begin{pmatrix}
1-\kappa(\pmb{\theta}) - \gamma_1(\pmb{\theta}) & -\gamma_2(\pmb{\theta}) \\
-\gamma_2(\pmb{\theta}) & 1-\kappa(\pmb{\theta}) + \gamma_1(\pmb{\theta})
\end{pmatrix}
\end{equation}
The solution of equation (\ref{geodesic3D}) via the multi--lens--plane algorithm yields a single realization of the WL fields $(\kappa,\pmb{\gamma})$. Multiple random realizations of these fields can be obtained by altering the lens system before ray--tracing using the randomization technique described in \citep{Petri16}.

\begin{figure*}
\includegraphics[scale=0.65]{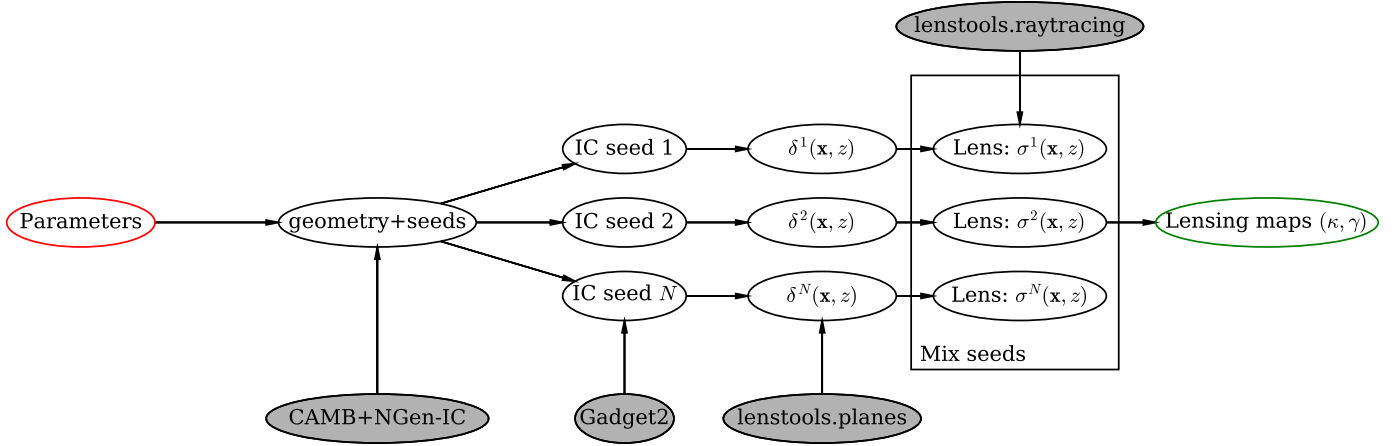}
\caption{Workflow of the \LT WL shear simulation operations, from the specifications of the cosmological parameters $\bb{p}$ to the finished image products. The diagram has to be read from left to right: the arrows originating from the grey nodes mean that the executable is run with the target node as an input. The produced output is passed down the pipeline on the right. Matter power spectra at high redshift are computed with \ttt{CAMB} and are used to generate the initial conditions for the $N$--body simulations (for which we use the \ttt{NGen-IC} add--on to the \ttt{Gadget2} code). These initial conditions, each with a different random seed, are then evolved in time with \ttt{Gadget2}. After the $N$--body snapshots are written to disk, \LT slices them into two dimensional lens planes. The slicing is done by an executable named \ttt{lenstools.planes}, which is a convenient wrapper for the operations implemented by the \ttt{NbodySnapshot} class. The ray--tracing operations are carried on by an executable named \ttt{lenstools.raytracing}, which conveniently wraps the operations implemented by the \ttt{RayTracer} class. From the final positions of the light rays the convergence $\kappa$ and the cosmic shear $\pmb{\gamma}$ can be inferred. Details on how to use the \ttt{lenstools.planes} and \ttt{lenstools.raytracing} executables, that come with \LT, can be found in the \LT documentation at the URL \url{http://lenstools.readthedocs.io/en/latest/pipeline.html}.}
\label{pipescheme}
\end{figure*}

The iterative solution of equations (\ref{geodesic2D})--(\ref{sheartensorproduct}) requires knowledge of the density fluctuation $\delta(\bb{x},z)$, from which the lens surface density $\sigma$ and gravitational potential $\phi$ can be inferred through equations (\ref{surfacedensity}),(\ref{poisson}). $\delta$ can be calculated running numerical simulations, such as $N$--body simulations \citep{Gadget2,HACC} or hydrodynamical simulations \citep{Flash}. In this paragraph we focus mainly on $N$--body simulations, in which the matter distribution in the universe is approximated as a set of $N_p$ particles of mass $M_p$, which move in their self--generated gravitational field. For this purpose we use the publicly available code \ttt{Gadget2}\citep{Gadget2}, although alternatives can be adopted (see \citep{HACC} for example). Once the $N$--body simulations are run, \LT provides a {\sc python} implementation of the multi--lens--plane algorithm \citep{RayTracingHartlap} described above, which takes care of projecting the density fluctuation on two dimensional lenses as in (\ref{surfacedensity}), solving the Poisson equation as in (\ref{poisson}), and computing the light ray deflections as in (\ref{geodesic2D})--(\ref{sheartensorproduct}). An overview of the pipeline operations, from the cosmological parameter specifications to the final shear map products, is outlined in Figure \ref{pipescheme}. We make the claim that the way the \LT code is organized makes it portable between different research groups that rely on different choices, rather than \ttt{Gadget2}, for running $N$--body simulations. The transition between snapshots and lens planes is handled by instances of the \ttt{NbodySnapshot} class, which implements the algorithms based on equations (\ref{poisson}). \ttt{NbodySnapshot} can be sub--classed to implement the necessary input routines from the $N$--body snapshots. \LT comes with one of such possibilities, the \ttt{Gadget2Snapshot} class, that handles input from snapshots in the \ttt{Gadget2} binary format. Other subclass types that allow input from other formats can be coded by the user with minimal effort.

\subsection{Pipeline code structure}
In this paragraph we describe how the \LT pipeline code is organized. The simulation products are placed in a directory tree structure designed for easy resource access. The directory tree is mirrored in two locations, a so called \textit{Home} location, which holds all the book--keeping information such as small data files (such as \ttt{CAMB} matter power spectra) and configuration files, and a \textit{Storage} location which holds the simulation products (\ttt{Gadget2} snapshots, lens planes and finished shear maps). The reason for this is that while the \textit{Home} location does not require much disk space, the \textit{Storage} location can reach disk sizes of several Terabytes. We found it convenient to keep the two locations separated to simplify sharing data among machines in a cluster, and help with portability issues among different clusters. 
A batch of simulations is handled by a single instance of a \ttt{SimulationBatch} object (which holds both the \textit{Home} and \textit{Storage} parts). The first level in the tree corresponds to a choice of values for cosmological parameters $\bb{p}$: each node on this level of the tree is an instance of the \ttt{SimulationModel} class. The second level in the tree specifies the size and resolution of the $N$--body box, namely the box size $L_b$ and the number of particles $N_p$: each node on the second level corresponds to a \ttt{SimulationCollection} object. Inside a simulation collection, we are free to choose different random realizations of the initial conditions, which will then be evolved in time by the $N$--body code. Each such realization lives on a node which is one level deeper in the tree, and is encoded in a \ttt{SimulationIC} object. The deepest level in the tree contains the two dimensional slices of the $N$--body simulation boxes. Each node on this level is an instance of the \ttt{SimulationPlanes} class. Once the lens planes are generated, the ray--tracing operations can be performed. In principle we can use lens planes that live under the same \ttt{SimulationCollection}, but belong to different \ttt{SimulationIC} nodes, to produce either single redshift shear images (each ensemble of images lives in a \ttt{SimulationMaps} object) or shear catalogs of $N_g$ sources, in the form of a table in which each of the $N_g$ rows is a tuple $(x_g,y_g,z_g,\gamma_{1,g},\gamma_{2,g})$. Each ensemble of catalogs corresponds to an instance of the \ttt{SimulationCatalog} class. Note that, because they combine lens planes with different initial random seed (see \citep{Petri16}), both \ttt{SimulationMaps} and \ttt{SimulationCatalog} nodes live on the same level of the directory tree, one level below \ttt{SimulationCollection}. An example on how to create a {\sc python} script to lay down such a directory tree is available in {\sc IPython} notebook format\footnote{\url{http://nbviewer.jupyter.org/github/apetri/LensTools/blob/master/notebooks/dirtree.ipynb}}. A comprehensive guide on how to deploy the \LT simulation pipeline on a computer cluster can be found in the code documentation \citep{lenstoolsdocs}.

\subsection{Performance}
We summarize the runtime and memory usage benchmarks of the \LT shear simulation pipeline. The tests were run on the XSEDE Stampede computer cluster\footnote{\url{https://portal.xsede.org/tacc-stampede}}. Table \ref{benchmarktable} shows a summary of the ray--tracing operations performed by \LT, indicating the complexity and runtime of each operation for a selected test case. At the lens plane generation stage we can clearly see that the two bottlenecks in the flow are the read operations from $N$--body snapshot files and the Poisson equation solving via FFT. The number of tasks $N_t$ used to read in a single snapshot can be optimized if the parallel input performance is ideal (i.e independent on $N_t$), as is an optimal value of $N_t(N_p,L_p)$ that minimizes the combined input, gridding and \ttt{MPI} communication operations, which have a combined complexity

\begin{equation}
t_{\rm in+grid+MPI} = A_1\frac{N_p}{N_t} + A_2L_p\log{N_t}
\end{equation} 
The optimal $N_t$ depends on the number of particles $N_p$ and the lens plane resolution (in pixels) $L_p$. In principle this bottleneck can be removed if the capability of generating lens planes is embedded into the $N$--body code, which could avoid saving the intermediate 3D snapshots to disk. \LT allows such a possibility by creating a channel of communication between \ttt{Gadget2} and the plane application using named pipes. For this option to be viable, \ttt{Gadget2} and the plane computation must run on the same node. 

The Poisson solver has a complexity of $O(L_p\log{L_p})$ which is dominated by FFT performance. Although we make use of the \ttt{numpy} FFT pack \citep{scipy} to perform such operations, other alternatives are also possible (such as FFTW \citep{FFTW05}). The \LT code modularity makes it very easy to switch between different implementations of the FFT algorithm, both coming from external libraries or coded up by the user. 
The bottleneck of the ray--tracing operations consists in the calculations of the ray deflections in equation (\ref{deflections}) and the distortion tensor products (\ref{sheartensorproduct}). Although the complexity of these operations is already optimal, improvements on the runtime can be made by using specialized libraries to handle matrix products. \ttt{numpy} can link to the most up--to--date version of specialized libraries such as \ttt{LAPACK} \citep{lapack} and \ttt{Intel MKL} \citep{intel-mkl}. Some versions of \ttt{numpy} even support automatic offload of linear algebra operations to \ttt{Intel XEON Phi} co--processors \citep{xeonphi}.     

\begin{figure}
\begin{center}
\includegraphics[scale=0.3]{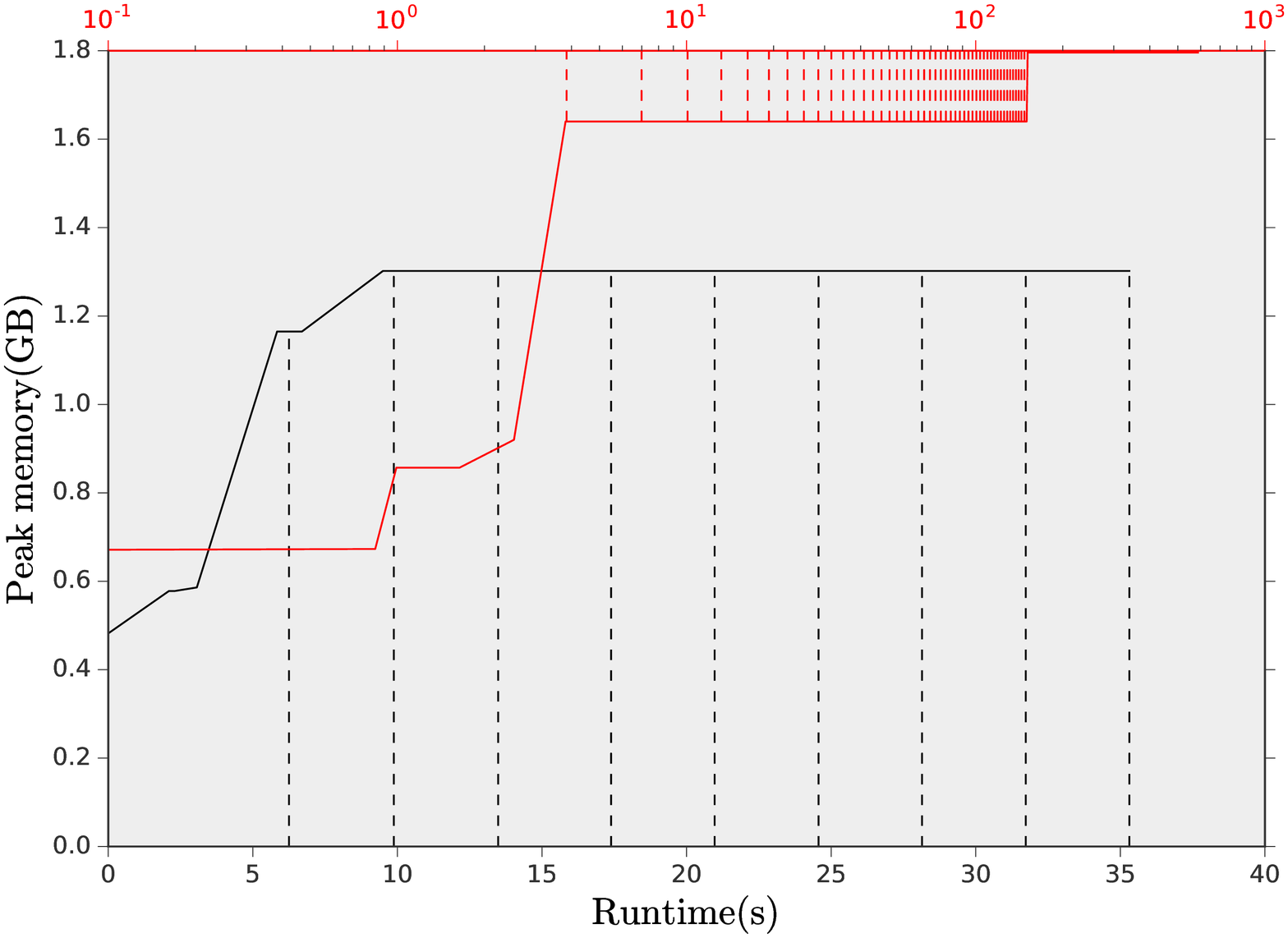}
\caption{Peak memory usage for the lens plane generation (black) and ray--tracing (red) as a function of runtime $t$ for the test case indicated in Table \ref{benchmarktable}. The vertical lines are drawn in correspondence of the completion of a lens plane calculation (black) and a lens crossing (red).}
\label{memoryfig}
\end{center}
\end{figure}

We tracked the memory usage of the lens plane and ray--tracing operations. This is an important step, since {\sc python} has some subtleties when dealing with large memory applications. Memory allocated by a {\sc python} process cannot be released manually as in {\sc C}, but is managed by the built--in garbage collector instead. Figure \ref{memoryfig} shows the peak memory usage during the lens plane generation and ray--tracing operations. We can see that, for the test case outlined in Table \ref{benchmarktable}, memory consumption stabilizes around 1.3\,GB per task for the lens planes and 1.8\,GB per task during the ray--tracing. These considerations make the \LT pipeline suitable for deployment on computer clusters with $\gtrsim$2\,GB memory per core, such as the one we used. These numbers refer to the test case described in Table \ref{benchmarktable}. Producing higher resolution lens planes and WL maps will in general require more memory. A rough estimate for the memory scaling with resolution can be made noting that the dominant contribution to the memory usage for the lens planes comes from the input from the $N$--body simulations and should hence scale with the number of particles $N_p$. For the ray--tracing, on the other hand, the dominant contribution to the memory usage comes from the lenses and should hence scale as the lens pixel resolution $L_p$, also because the resolution of the WL maps $N_r$ needs to be smaller than $L_p$.     

\begin{table*}
\begin{center}
\begin{tabular}{l|c|c|c}
\toprule
{Step} &            Complexity &            Test case &           Runtime \\ \hline \hline
\midrule
\multicolumn{4}{c}{\textbf{Lens generation}} \\ \hline
Snapshot input & $O(N_p/N_t)$  & $N_p=512^3$, $N_t=16$  & 2.10\,s  \\
Gridding        & $O(N_p/N_t)$   & $N_p=512^3$, $N_t=16$  & 0.20\,s \\
\ttt{MPI} Communication  & $O(L_p\log{N_t})$   & $N_t=16$, $L_p=4096^2$  & 0.76\,s   \\
Poisson solver (FFT)           & $O(L_p\log{L_p})$ & $L_p=4096^2$  &  2.78\,s    \\
Lens output           & $O(L_p)$ & $L_p=4096^2$   & 0.04\,s  \\ \hline \hline

\multicolumn{4}{c}{\textbf{Ray tracing}} \\ \hline
Lens input &  $O(L_p)$ & $L_p=4096^2$ & 0.32\,s \\
Random lens shift &  $O(L_p)$ & $L_p=4096^2$ & 0.15\,s \\
Deflection calculation        &  $O(N_r)$ & $N_r=2048^2$   & 1.54\,s  \\
Shear tensor product               &  $O(N_r)$ & $N_r=2048^2$   &  1.29\,s \\ \hline \hline

\bottomrule
\end{tabular}
\caption{Summary of the ray--tracing benchmarks: each $N$--body snapshot is divided in $N_t$ files, which are read in parallel and contain a total of $N_p$ particles (perfect input performance is assumed in the complexity analysis). After the gridding procedure (\ref{surfacedensity}) is performed by each task, the total sufrace density (computed for a plane of $L_p$ pixels) is collected by the master task, which then proceeds in solving the Poission equation (\ref{poisson}) via Fast Fourier Transforms and saves the output to disk. In a subsequent step, the lens potential files are read from disk, and the geodesic equations (\ref{geodesic2D}) are solved for $N_r$ different starting positions $\pmb{\theta}$ that allow to reconstruct the shear and convergence fields $\pmb{\gamma},\kappa$. The numbers refer to tests conducted on the XSEDE Stampede cluster. Parallel operations are implemented with \ttt{mpi4py} \citep{mpi4py}, a {\sc python} wrapper of the \ttt{MPI} library \citep{MPI}.}
\label{benchmarktable}
\end{center}
\end{table*}

\section{Image analysis}
\begin{figure*}
\begin{center}
\includegraphics[scale=0.35]{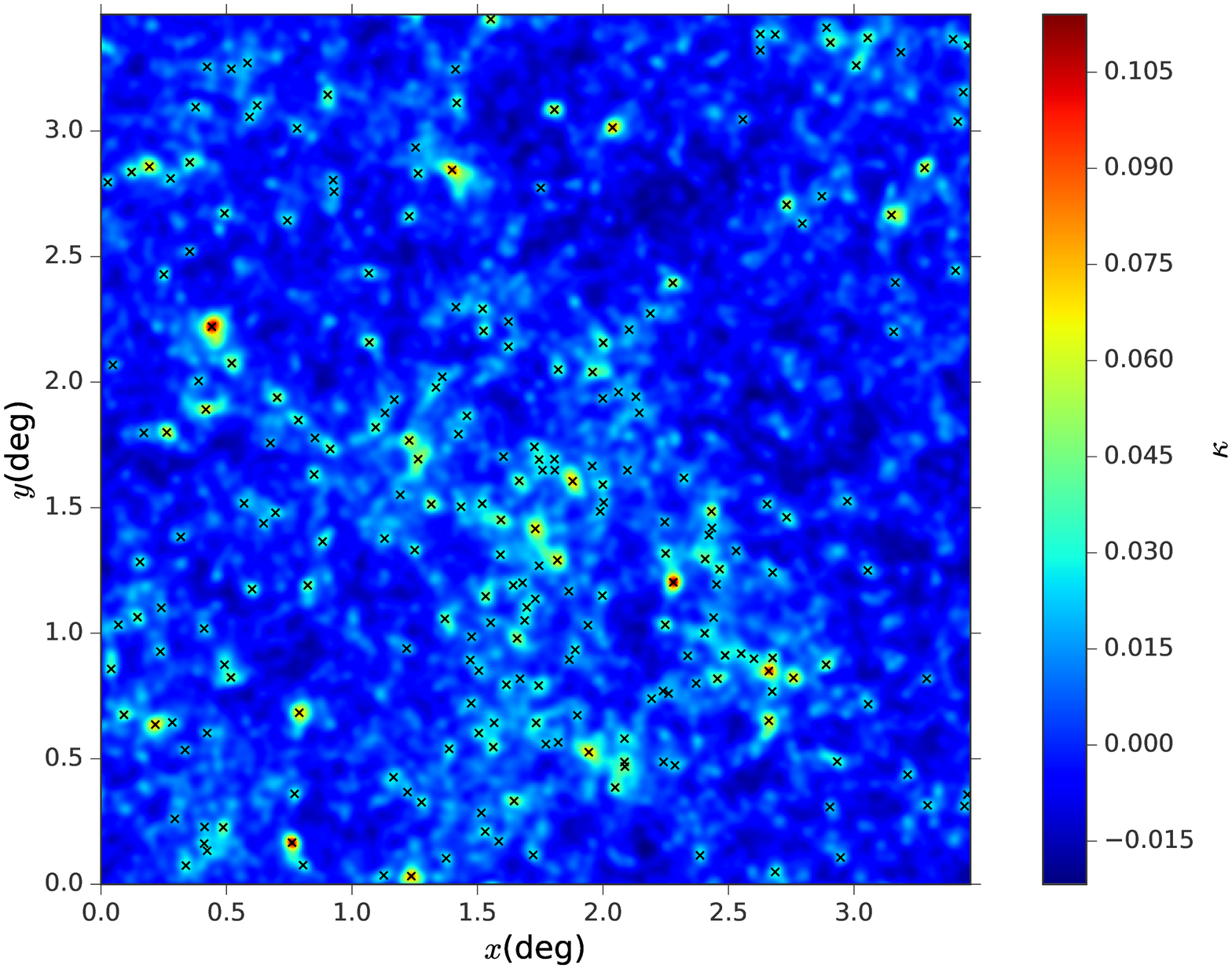}
\includegraphics[scale=0.3]{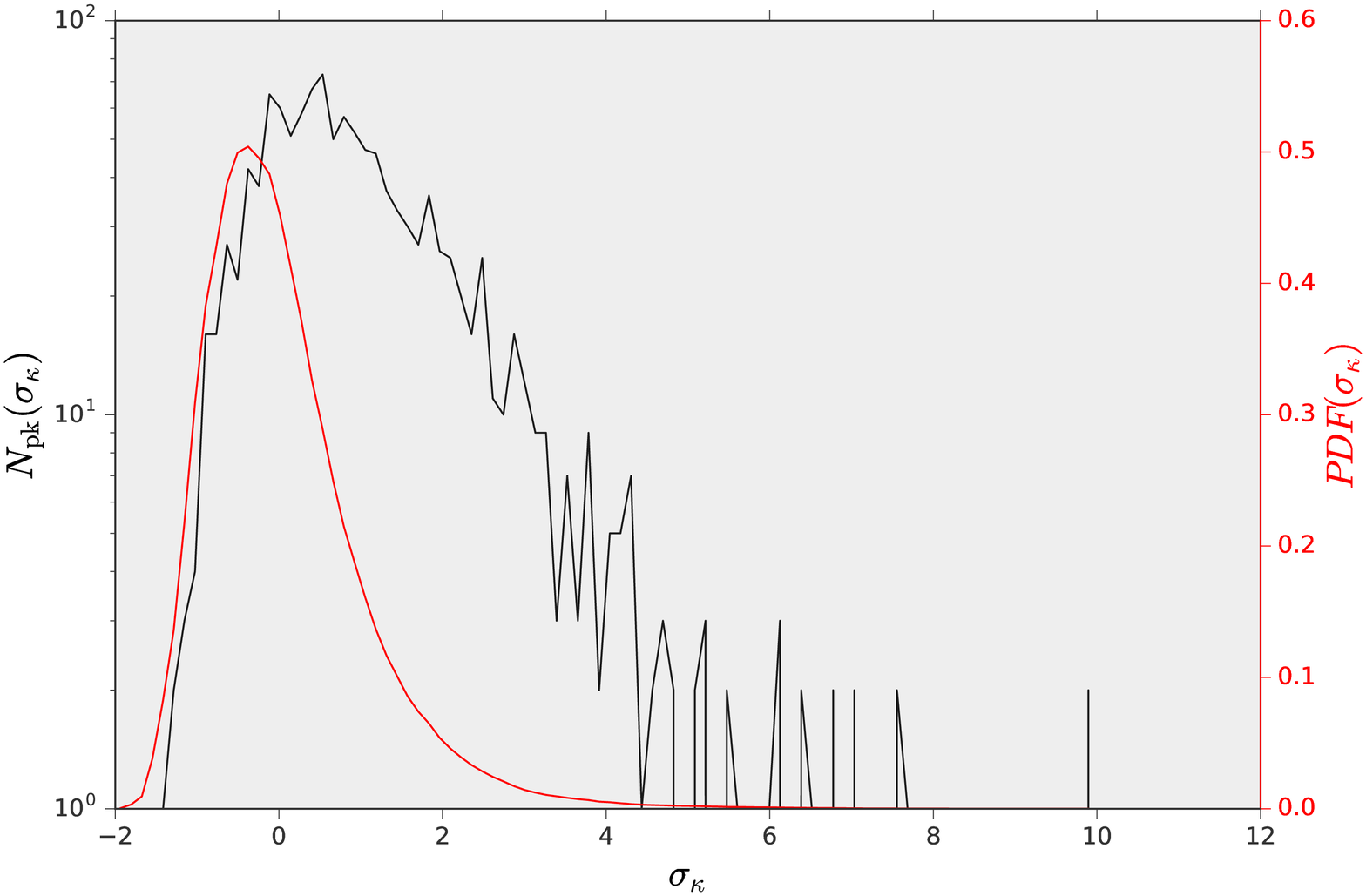}
\caption{One of the convergence maps produced with the \LT shear simulation pipeline. The map has been generated assuming a uniform background source distribution at $z_s=2$ and has an angular size of $3.5^\circ$ and a resolution of 2048 pixels per side, which correspond to a pixel resolution of $0.1^\prime$. The black crosses in the left panel identify local maxima (\textit{peaks}) in the $\kappa$ field with a significance of at least 2$\sigma$. The right panel shows the PDF of the $\kappa$ field (red) and its peak histogram (black). The code to produce this figure is available in {\sc IPython} notebook format at \url{http://nbviewer.jupyter.org/github/apetri/LensTools/blob/master/notebooks/image.ipynb}.}
\label{convergencefig}
\end{center}
\end{figure*}

In this section we describe some useful routines that \LT provides for analyzing simulated convergence and shear fields. The \ttt{ConvergenceMap} and \ttt{ShearMap} classes implement several operations that can be performed on two dimensional $\kappa,\pmb{\gamma}$ images (for a complete list look at the \LT documentation \citep{lenstoolsdocs}). Both classes allow flexible I/O formats from files through the \ttt{load} method (the FITS format \citep{cfitsio} is a popular choice, but not the only one possible; user custom format can be easily dealt within \LT). Efficient routines are available for smoothing the maps with Gaussian kernels (optimal smoothing complexity--wise can be performed via FFT for kernel sizes bigger than $\sim 10$ pixel, otherwise real--space techniques are preferrable. \LT allows for both possibilities.), measuring the pixel PDF, counting the local maxima and measuring their position, measuring topological descriptors such as Minkowski Functionals \citep{MatsubaraMink}. An example of such operations is shown in Figure \ref{convergencefig}.  

\begin{figure}
\includegraphics[scale=0.3]{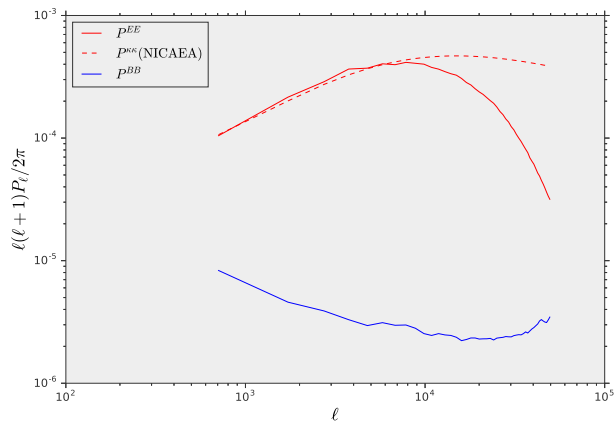}
\caption{$E$ and $B$ mode power spectra of one of the shear maps generated with the \LT simulation pipeline. We show the $E$--mode power spectrum $P^{EE}$ (red) and the $B$--mode power spectrum (blue), computed using equation (\ref{ebmodeeqs}) We also show an analytical prediction of the $\kappa$ power spectrum obtained with the public code \ttt{Nicaea} (dashed red line). There is a discrepancy between the simulations and the analytical results at high $\ell$ which has to do with the finite resolution of the $N$-body simulations and of the WL maps. This finite resolution causes the $\kappa$ fields to lack power on small scales.}
\label{ebmodefig}
\end{figure}

In addition to real space statistics such as the ones outlined before, \LT provides access to quadratic Fourier statistics such as power spectra. The convergence angular power spectrum $P^{\kappa\kappa}(\ell)$ is defined as  
\begin{equation}
\label{kappapsdefinition}
\langle\tilde{\kappa}(\pmb{\ell})\tilde{\kappa}(\pmb{\ell}^\prime)\rangle = (2\pi)^2\delta_D(\pmb{\ell}+\pmb{\ell}^\prime)P^{\kappa\kappa}(\ell)
\end{equation}
where we indicate the Fourier transform of $\kappa(\pmb{\theta})$ as $\tilde{\kappa}(\pmb{\ell})$. The shear field can be decomposed into its $E$ and $B$ components according to 

\begin{equation}
\label{ebmodeeqs}
\begin{matrix}
E(\pmb{\ell}) = \frac{(\ell_x^2-\ell_y^2)\tilde{\gamma}^1(\pmb{\ell})+2\ell_x\ell_y\tilde{\gamma}^2(\pmb{\ell})}{\ell_x^2+\ell_y^2}  \\ \\
B(\pmb{\ell}) = \frac{-2\ell_x\ell_y\tilde{\gamma}^1(\pmb{\ell})+(\ell_x^2-\ell_y^2)\tilde{\gamma}^2(\pmb{\ell})}{\ell_x^2+\ell_y^2}
\end{matrix}
\end{equation}
Because of the nature of the density perturbations that cause background source lensing, the $E$ component dominates in the weak lensing limit, because $E=O(\phi)$ and $B=O(\phi^2)$. The shear $E$--mode is the convergence $\tilde{\kappa}$ (in Fourier space), hence $P^{EE}(\ell)=P^{\kappa\kappa}(\ell)$. Figure \ref{ebmodefig} shows the shear $E$ and $B$ modes power spectra measured from one realization of the shear field.   

\section{Cosmology constraints}

\subsection{Formalism}
In this section we outline the basic routines that \LT provides for constraining cosmology. Let $\bb{p}$ be the parameters to constrain (for example $\bb{p}=(\Omega_m,w,\sigma_8)$) and let $\bb{d}$ be a WL feature that contains information on $\bb{p}$. An example of $\bb{d}$ can be the $\kappa$ power spectrum $P^{\kappa\kappa}(\ell)$ defined in (\ref{kappapsdefinition}). Given a feature measurement $\bbh{d}$, we are interested in calculating the likelihood $\mathcal{L}(\bb{p}\vert\bbh{d})$ of the parameters, given the measurement. Using the Bayes theorem we can express the parameter likelihood as 
\begin{equation}
\label{parameterlikelihood}
\mathcal{L}(\bb{p}\vert\bbh{d}) = \mathcal{N}_\mathcal{L}\mathcal{L}(\bbh{d}\vert\bb{p})\Pi(\bb{p})
\end{equation} 
where $\Pi(\bb{p})$ is the prior on the parameters, $\mathcal{N}_\mathcal{L}$ is a $\bb{p}$--independent normalization constant and $\mathcal{L}(\bbh{d}\vert\bb{p})$ is the feature likelihood. A popular choice for the feature likelihood is a normal distribution with mean $\bb{d}(\bb{p})$ and covariance $\bb{C}$, $\mathcal{L}(\bbh{d}\vert\bb{p}) = \exp{\left[-\chi^2\left(\bbh{d}\vert\bb{d}(\bb{p}),\bb{C}\right)/2\right]}$ with

\begin{equation}
\label{chi2definition}
\chi^2\left(\bbh{d}\vert\bb{d}(\bb{p}),\bb{C}\right) = (\bbh{d}-\bb{d}(\bb{p}))^T\bb{C}^{-1}(\bbh{d}-\bb{d}(\bb{p}))
\end{equation}
\LT provides efficient routines for computing (\ref{chi2definition}) at arbitrary points $\bb{p}$ in parameter space. We argue that efficient evaluation of $\chi^2$ is not only useful when data likelihoods are Gaussian, but also gives access to more advanced likelihood sampling methods such as Approximate Bayesian Computation \citep{cosmoabc}.  
Running the simulation pipeline described in \S~\ref{shearsim} for a variety of cosmological models $\{\bb{p}_i\}$, $i=1...N_M$, gives access to a discrete set of features $\{\bb{d}(\bb{p}_i)\}$ evaluated at $\bb{p}_i$. We are able to evaluate equation (\ref{chi2definition}) at an arbitrary point $\bb{p}$ in parameter space using a Radial Basis Function (RBF) interpolation scheme 
\begin{equation}
\label{rbf}
d^{RBF}_i(\bb{p}) = \sum_{j=1}^{N_M}\lambda_{ij} f\left(\left\vert\bb{p}-\bb{p}_j\right\vert;R\right)
\end{equation} 
where $f$ is an isotropic smoothing kernel of scale $R$ \footnote{For the examples shown in this work we chose a multiquadric kernel $f(x;R)=\sqrt{1+\frac{x^2}{R^2}}$ with $R$ chosen as the mean distance between the simulated points $\bb{p}_k$} and the weights $\lambda_{ij}$ can be determined from the known simulated features $\bb{d}(\bb{p}_j)$ with a matrix inversion
\begin{equation}
\lambda_{ij} = \sum_{k=1}^{N_M}d_i(\bb{p}_k)\left[f\left(\left\vert\bb{p}_k-\bb{p}_j\right\vert;R\right)\right]^{-1}
\end{equation}
Once the weights are determined from the simulated features, equation (\ref{rbf}) allows for a fast vectorized evaluation of the parameter likelihood from equations (\ref{parameterlikelihood}),(\ref{chi2definition}) once an assumption for the ($\bb{p}$--independent) covariance matrix $\bb{C}$ is made. A possible choice is estimating the covariance $\bb{C}$ from the shear simulations themselves \footnote{In the case in which the covariance matrix is estimated from simulations, the estimator for its inverse $\bbh{C}^{-1}$ is biased \citep{Taylor12}. \LT sampling routines use the unbiased estimator $\frac{N_r-N_b-2}{N_r-1}\bbh{C}^{-1}$ where $N_r$ is the number of realizations used to estimate the $N_b\times N_b$ feature covariance}. Having access to an efficient routine for computing the parameter likelihood $\mathcal{L}(\bb{p}\vert\bbh{d})$ allow access to a variety of parameter sampling techniques. In this section we give examples of three different parameter sampling techniques supported in \LT:
\begin{enumerate}

\item Likelihood Grid Evaluation: if the dimensionality of the parameter space is not too big, the parameter likelihood in equation (\ref{parameterlikelihood}) can be evaluated on a regularly spaced grid of points. Since each point can be treated independently, this procedure is easily parallelizable. \LT provides a parallel implementation of the likelihood grid evaluation based on the \ttt{MPI} protocol \citep{MPI}. Access to the values of the likelihood on a regularly spaced grid makes the determination of confidence intervals straightforward. Such an approach has been used before in the literature \citep{cfhtpeaks,cfhtmink}.    

\item MCMC sampling of the parameter space: the efficient \LT likelihood evaluation routines are specifically designed to be compatible with widely used {\sc python} packages such as \ttt{emcee} \citep{emcee} and \ttt{pymc} \citep{pymc} that specialize in generating parameter samples with the Markov Chain Monte Carlo (MCMC) technique \citep{mcmc}. \ttt{emcee} supports parallel MCMC sampling via \ttt{mpi4py}.    

\item Fisher Matrix approximation: for the sake of simplicity, sometimes it is convenient to approximate the parameter likelihood as a Gaussian centered around its maximum. If we have a reasonable guess for the likelihood peak location $\bb{p}_0$ (in the case where the likelihood is single--modal), the parameter covariance matrix can be approximated as 
\begin{equation}
{\rm Cov}(p_\alpha,p_\beta) = -\left(\left.\frac{\partial\ln\mathcal{L}(\bb{p}\vert\bbh{d})}{\partial p_\alpha\partial p_\beta}\right\vert_{\bb{p}_0}\right)^{-1}
\end{equation}
The partial derivatives of the likelihood with respect to the parameters are easy to evaluate with finite differences. 
\end{enumerate}

\subsection{Code structure}

\begin{figure}
\begin{center}
\includegraphics[scale=0.4]{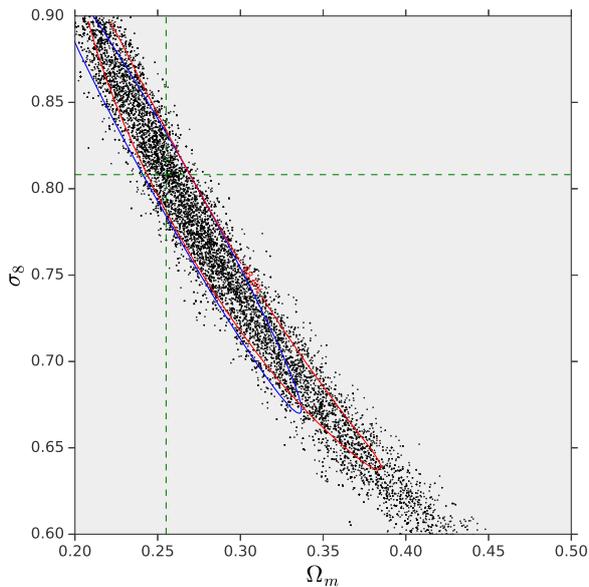}
\caption{Cosmological constraints based on the sampling of the parameter space $(\Omega_m,\sigma_8)$ with three different techniques, assuming a flat parameter prior $\Pi(\bb{p})$. We use a Likelihood Grid Evaluation of parameter likelihood (red), a Fisher matrix approximation (blue) and a MCMC sampling using the \ttt{emcee} package (black dots). The parameter space considered is the doublet $(\Omega_m,\sigma_8)$, and the feature used is the $\kappa$ power spectrum $P^{\kappa\kappa}(\ell)$. The $1\sigma$ confidence level contours are shown. The RBF interpolation is based on $N_M=100$ simulated models and the feature covariance matrix $\bb{C}$ has been estimated from $N_r=1024$ random realizations of the feature vector in a fiducial model with $(\Omega_m,\sigma_8)=(0.26,0.8)$. The code to produce this figure is available in {\sc IPython} notebook format at \url{http://nbviewer.jupyter.org/github/apetri/LensTools/blob/master/notebooks/sampling.ipynb}.}
\label{samplingfig}
\end{center}
\end{figure}

We give a brief overview of the \LT object types that handle parameter constraints operations. Feature emulators in \LT are row--oriented data structures, in which each of the $N_M$ rows contains a tuple of cosmological (and/or nuisance) parameters $\bb{p}_i$ and the simulated feature at $\bb{p}_i$. The base class that handles row--oriented data in \LT is the \ttt{Ensemble} class. \ttt{Ensemble} inherits from \ttt{pandas.DataFrame} \citep{pandas} and provides additional routines for feature measurement from simulations (both serial and parallel in the number of $\kappa$ maps with \ttt{mpi4py} \citep{mpi4py}), Principal Component Analysis, statistics bootstrapping. As a sub--class of \ttt{pandas.DataFrame}, \ttt{Ensemble} supports I/O and queries from local and remote SQL databases. Parameter space sampling operations in \LT are handled by instances of the \ttt{Emulator} class, which inherits from \ttt{Ensemble} and provides access to vectorized RBF feature interpolation between parameters (which can be used to build emulators), $\chi^2$ evaluation, and the parameter sampling techniques outlined above. Examples of parameter sampling routines are shown in Figure \ref{samplingfig}.          

\section{Conclusion}
In this work we presented the \LT computing package, which is a collection of tools targeted to theoretical studies of WL. The package includes a WL shear simulation pipeline and is complemented with a variety of image analysis tools and parameter sampling routines. The simulation pipeline combines different existing codes to simulate cosmological volumes (\ttt{CAMB},\ttt{Gadget2}) and provides a {\sc python} implementation of the multi--lens--plane algorithm. \LT is flexible in terms of snapshot file formats, making the use of different $N$--body simulation codes (see \citep{HACC} for example) possible with minimal additional efforts. This makes \LT portable between different research groups in the WL community. \LT makes use of the \ttt{numpy} array as its primary data structure for numerical calculations, making it very convenient to combine with popular algorithmic packages for data selection (\ttt{pandas} \citep{pandas}), astronomical tools (\ttt{astropy} \citep{astropy}), MCMC sampling (\ttt{pymc} \citep{pymc}, \ttt{emcee} \citep{emcee}), advanced statistical analysis and machine learning (\ttt{scikit-learn} \citep{scikit-learn}). Because of these reasons, we believe that \LT will become a valuable asset to the WL community, in particular to groups that already have experience with $N$--body simulations and want to study Weak Lensing.


\section*{Acknowledgements}
We thank Zolt\`an Haiman, Jia Liu, Jose M. Zorrilla and Morgan May for invaluable support and useful discussions. The simulations in this work were performed at the NSF Extreme Science and Engineering Discovery Environment (XSEDE), supported by grant number ACI-1053575, at the Yeti computing cluster at Columbia University, and at the New York Center for Computational Sciences, a cooperative effort between Brookhaven National Laboratory and Stony Brook University, supported in part by the State of New York. This work was supported in part by the U.S. Department of Energy under Contract Nos. DE-AC02-98CH10886 and DESC0012704, and by the NSF Grant No. AST-1210877 (to Z.H.) and by the Research Opportunities and Approaches to Data Science (ROADS) program at the Institute for Data Sciences and Engineering at Columbia University. 

\bibliography{ref}

\label{lastpage}
\end{document}